\documentclass[tradiabstract,letter]{aa}  

\usepackage{graphicx}
\usepackage{txfonts}
\usepackage{subcaption}         
\usepackage{lscape}             
\usepackage{placeins}           
\usepackage{appendix}
\usepackage{hyperref}
\usepackage{amssymb}

\begin{document}
\newcommand{\Ho}{\mbox{$H_0$}}
\newcommand{\ang}{\mbox{{\rm \AA}}}
\newcommand{\abs}[1]{\left| #1 \right|} 
\newcommand{\avg}[1]{\left\langle #1 \right\rangle} 
\newcommand{\kms}{\ensuremath{{\rm km\,s^{-1}}}}
\newcommand{\zabs}{\ensuremath{z_{\rm abs}}}
\newcommand{\zem}{\ensuremath{z_{\rm quasar}}}
\newcommand{\cmsq}{\ensuremath{{\rm cm}^{-2}}}
\newcommand{\ergs}{\ensuremath{{\rm erg\,s^{-1}}}}
\newcommand{\ergsa}{\ensuremath{{\rm erg\,s^{-1}\,{\AA}^{-1}}}}
\newcommand{\ergscm}{\ensuremath{{\rm erg\,s^{-1}\,cm^{-2}}}}
\newcommand{\ergscma}{\ensuremath{{\rm erg\,s^{-1}\,cm^{-2}\,{\AA}^{-1}}}}
\newcommand{\msyr}{\ensuremath{{\rm M_{\rm \odot}\,yr^{-1}}}}
\newcommand{\nhi}{n_{\rm HI}}
\newcommand{\fhi}{\ensuremath{f_{\rm HI}(N,\chi)}}
\newcommand{\refs}{{\bf (refs!)}}
\newcommand{\Av}{\ensuremath{A_V}}
\newcommand{\lya}{Ly-$\alpha$}
\newcommand{\lyb}{Ly-$\beta$}
\newcommand{\lyg}{Ly-$\gamma$}

\newcommand{\Hb}{H-$\beta$}
\newcommand{\OVI}{\ion{O}{vi}}
\newcommand{\OIII}{\ion{O}{iii}}
\newcommand{\OII}{\ion{O}{ii}}
\newcommand{\OI}{\ion{O}{i}}
\newcommand{\HI}{\ion{H}{i}}
\newcommand{\HeII}{\ion{He}{ii}}
\newcommand{\HH}{\ensuremath{{\rm H}_2}}
\newcommand{\SII}{\ion{S}{ii}}
\newcommand{\SiIII}{\ion{Si}{iii}}
\newcommand{\SiIV}{\ion{Si}{iv}}
\newcommand{\SiII}{\ion{Si}{ii}}
\newcommand{\AlIII}{\ion{Al}{iii}}
\newcommand{\AlII}{\ion{Al}{ii}}
\newcommand{\ArI}{\ion{Ar}{i}}
\newcommand{\FeII}{\ion{Fe}{ii}}
\newcommand{\ZnII}{\ion{Zn}{ii}}
\newcommand{\CrII}{\ion{Cr}{ii}}
\newcommand{\MnII}{\ion{Mn}{ii}}
\newcommand{\MgII}{\ion{Mg}{ii}}
\newcommand{\MgI}{\ion{Mg}{i}}
\newcommand{\NiII}{\ion{Ni}{ii}}
\newcommand{\NV}{\ion{N}{v}}
\newcommand{\CIV}{\ion{C}{iv}}
\newcommand{\CIII}{\ion{C}{iii}}
\newcommand{\CII}{\ion{C}{ii}}
\newcommand{\CI}{\ion{C}{i}}
\newcommand{\CaII}{\ion{Ca}{ii}}
\newcommand{\TiII}{\ion{Ti}{ii}}
\newcommand{\Jlong}{SDSS\,J080023.02+305101.22}
\newcommand{\J}{J\,0800+3051}

\newcommand{\iap}{Institut d'Astrophysique de Paris, CNRS-SU, UMR\,7095, 98bis bd Arago, 75014 Paris, France -- \email{noterdaeme@iap.fr}\label{iap}}

\newcommand{\ioffe}{Ioffe Institute, {Polyteknicheskaya 26}, 194021 Saint-Petersburg, Russia \label{ioffe}}

\newcommand{\cral}{Universit{\' e} Lyon 1, ENS de Lyon, CNRS, CRAL, UMR 5574, Saint-Genis-Laval, France \label{cral}}
    
\newcommand{\fcla}{French-Chilean Laboratory for Astronomy, IRL 3386, CNRS and U. de Chile, Casilla 36-D, Santiago, Chile \label{fcla}
             }


   \title{
   Molecular gas hidden in plain sight in the early Universe}
   
%

   \author{Pasquier Noterdaeme\inst{\ref{iap}} 
   \and 
   Jens-Kristian Krogager\inst{\ref{fcla},\ref{cral}}
   \and 
   Sergei Balashev\inst{\ref{ioffe}}
        }

   \institute{\iap \and \fcla \and \cral \and \ioffe
   }

   \date{\today}

  \abstract{
We report the discovery of an extreme intervening molecular absorber at $\zabs$\,=\,4.1  towards the $z$\,=\,4.7 quasar \Jlong\ revealed through strong \HH\ 
 absorption that had remained unnoticed in archival data for about two decades.
The system, 
which we analysed  
with new VLT/X-shooter observations, has a metallicity of about one-fifth the solar value, in line with the moderate dust reddening $\Av$\,$\sim$\,0.06\,mag. More strikingly, it exhibits the highest molecular fraction, $f_{\HH}= 2N(\HH)/(2N(\HH)+N(\HI)) \approx 60$\,\%, measured at $z>0$ from direct determinations of both atomic and molecular hydrogen column densities. Notably, this fraction is comparable to the very highest values in the Local Group. 
The inferred $f_{\HH}$ still represents a conservative lower limit to the local 
molecular fraction since the \HI\ absorption very likely includes atomic gas unrelated to the actual molecular component, as indicated by the wide ($\Delta v\sim 500$~\kms) multi-component low-ionisation metal profile. 

The detection of such a system at early cosmic times is remarkable given the limited number of quasar spectra probing $z>4$. It suggests an unexpectedly high incidence of \HH, with important implications for the evolution of molecular gas. This high incidence may be driven by high average densities and enhanced turbulence at those redshifts. %
At the same time, our results also highlight possible observational biases, both in quasar selection and in recognising strong \HH\ absorption, suggesting that a significant fraction of molecular gas may remain undetected, in particular near the peak of star formation.

  {The present system offers a unique benchmark for developing efficient detection algorithms. Further progress will benefit from colour-independent quasar surveys, while constraining the physical conditions and environments of such extreme absorbers will require observations on future extremely large telescopes.}
  }

   \keywords{Quasars: absorption lines -- ISM: molecules}

   \maketitle

\section{Introduction}
The central role of molecular gas in galaxy evolution is no longer in question, as it constitutes the immediate fuel for star formation. What remains less constrained, however, is where and under which  
conditions the bulk of molecular gas resides, especially at high redshift.
In fact, the dominant constituent of molecular gas, H$_2$, is notoriously difficult to detect in emission, so  indirect tracers such as CO are widely used instead. While absorption spectroscopy provides a direct, distance-insensitive probe of H$_2$ through the electronic Lyman and Werner bands,  
the small cross-section of cold, dense molecular clouds implies that intercepting such gas beyond the Local Group along random sightlines, i.e. at $z>0$ towards quasars, is intrinsically rare.

Despite its fundamental importance and significant observational progress \citep[][based on average H$_2$ absorption signal]{Balashev2018}, 
the column-density distribution of H$_2$ remains poorly constrained at its high end. The most molecular-rich absorption systems are not only intrinsically the rarest, but they are also suspected to be systematically missed: H$_2$ absorbers are commonly associated with dust reddening of the background quasar, particularly at high metallicity \citep[e.g.][]{Noterdaeme2017,Balashev2019}. At high column densities, the many H$_2$ lines also blend together and absorb most of the far-UV flux \citep{Noterdaeme2015b}. 
As a result, in addition to strengthening potential biases in the quasar photometric selection \citep{Fall1993,Krogager2020}, the strongest H$_2$ systems could in fact also be difficult to recognise in the quasar spectra, especially at high redshift, where the overall transmission of the intergalactic medium is already low.

Here we report the discovery of an extreme intervening H$_2$ absorber associated with the $z=4.096$ damped \lya\ system (DLA) towards the bright $z=4.67$ quasar \Jlong\ (hereafter \J). This is only the third H$_2$ absorption system detected at $z>4$ \citep{Ledoux2006, Noterdaeme2026}, but, most importantly, it has the highest molecular fraction directly measured (i.e. from H$_2$ and H\,{\sc i} lines) at $z>0$. 
Finally, and somewhat paradoxically, this quasar and the intervening DLA have been known for about two decades, but the strong H$_2$ system remained unrecognised until now.
We present {data of \J} 
from different surveys in Sect.~\ref{s:data} along with new observations collected with the X-Shooter spectrograph on the Very Large Telescope (VLT) {of the European Southern Observatory (ESO)}. We present our analysis and properties of the system in Sect.~\ref{s:meas} and discuss our results and conclude in Sect.~\ref{s:discussion}.

\section{Survey data and new observations\label{s:data}}

\subsection{Public survey data}

The object \J\ was first targeted for spectroscopic observations in early 2003 by the Sloan Digital Sky Survey \citep[SDSS;][with resolution power of $R\sim2000$]{York2000} as a high-redshift quasar candidate \citep{Richards2002}. Its formal identification as a quasar was then reported by \citet{Schneider2005} based on this spectrum. \citet{Prochaska2005sdss} subsequently detected the $z=4.1$ DLA using an automated procedure followed by visual inspection and fit of the \HI\ line, reporting $\log N(\HI)/\cmsq=20.8$. The quasar was re-observed by the SDSS-III in 2010 in the course of the Baryon Oscillation Spectroscopic Survey \citep{boss}. 

The high redshift and luminosity of the quasar motivated several follow-up studies, both to characterise the quasar itself \citep[e.g.][]{Trakhtenbrot2011} and to investigate intervening diffuse gas at early cosmic times. 
The aim of the Giant Gemini GMOS survey \citep[][with $R\sim 900$ but high S/N]{Worseck2014} was to constrain the mean free path of ionising photons at high redshift and identify DLAs \citep{Crighton2015} as well as Lyman-limit systems \citep{Crighton2019}, the latter based on the detection of sharp flux discontinuities bluewards of the Lyman limit.
These studies also involved visual inspections of the quasar spectrum, including focusing on the DLA, but they did not report the presence of strong molecular absorption. Finally, this system did not pass the threshold confidence cuts in the first automated search for H$_2$ absorption in DLAs either \citep{Balashev2014}.

\subsection{New observations: VLT/X-shooter and 4MOST}

Following a serendipitous identification of the system, we observed \J\ twice with X-shooter \citep{Vernet2011} through Director's Discretionary Time on January 25, 2026, under Prog ID 116.2AQ1 (PI Noterdaeme). Each observation consisted of a $\sim$ 3000\,s exposure for 
the UVB and VIS arms and 5$\times$600\,s exposures for the near-infrared, with slit widths of respectively 1.0, 0.9, and 0.9". The observations were performed in dark time under good conditions (clear sky with 0.7" seeing), though at relatively high airmasses (1.7-2) given the high declination of the target. The airmass-corrected seeing being slightly larger than the slit widths, the nominal spectral resolutions were obtained with $R=5400$, 8900, and 5600 for the UVB, VIS, and near-infrared, respectively.
The data were reduced using the ESO pipeline v3.5.3 to produce the 1D spectra for each arm, which we combined using an inverse-variance weighting. Absorption from the Earth's atmosphere was modelled for each exposure using Molecfit \citep{Smette2015}, and the corresponding transmissions were combined in the same way as the science data to produce an effective atmospheric transmission model. 

The quasar was also serendipitously observed for 1 hour during commissioning of 4MOST \citep{deJong2019} in December 2025 in order to check the effectiveness of observations at high airmass and test the performance of the redshift pipeline. 
The 4MOST data, covering 3700 to 9500~\AA\ at $R \approx 5000$, were processed using the official pipeline v0.9 (from date December 22, 2025), with daytime calibrations obtained the morning after. 

\section{Properties of the molecular absorption system \label{s:meas}}
\begin{figure*}
\centering
    \includegraphics[trim={1cm 0.1cm 1.2cm 0cm},width=0.98\hsize,clip=]{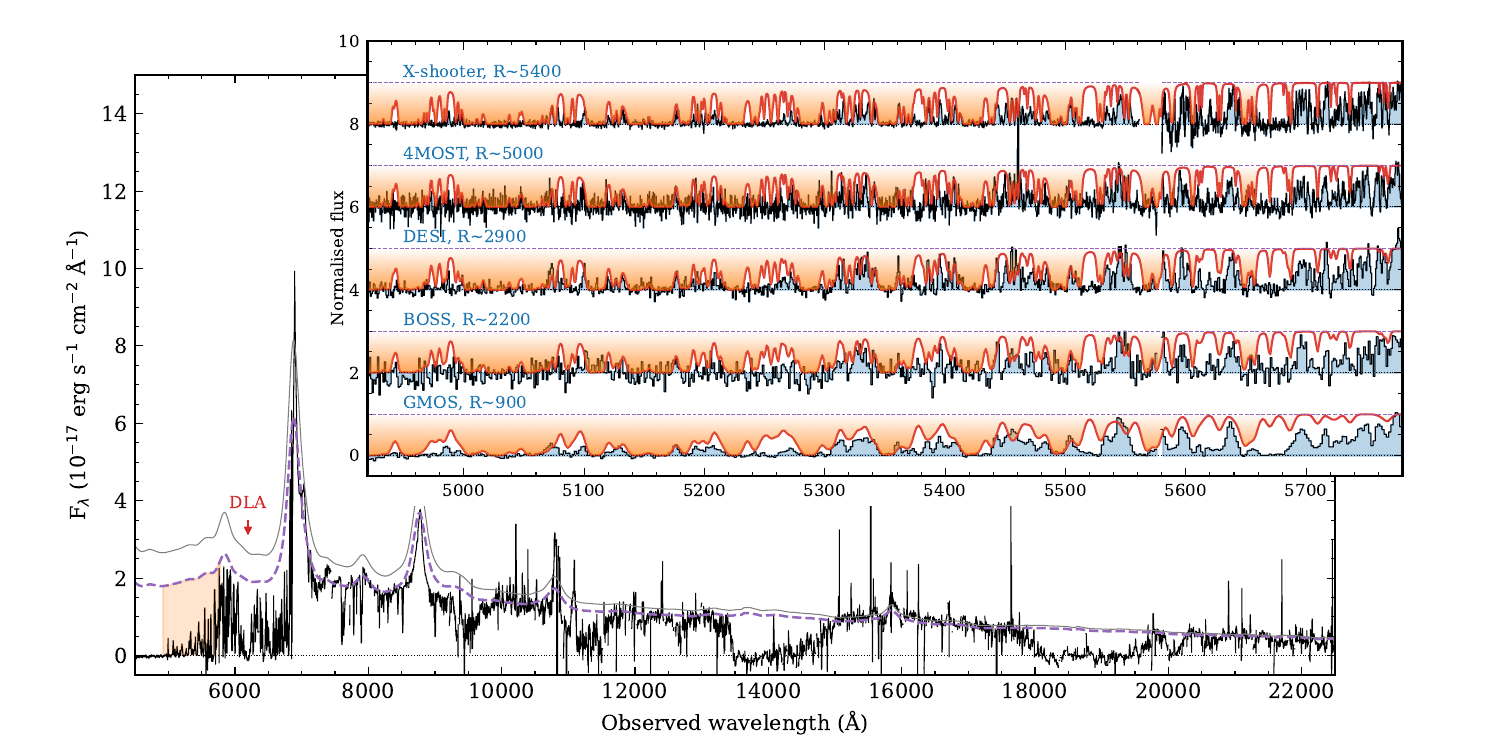}
    \caption{X-shooter spectrum of the $z=4.67$ quasar \J\ (black, smoothed for visual purpose) featuring a DLA at $z=4.096$ (red arrow).  
    The composite quasar spectrum from \citet[][]{Selsing2016} is shown in grey, and the same composite reddened by 
    \Av=0.06~mag at the DLA redshift is shown as a purple dashed line.
    The significant decrease in transmission at short wavelengths (orange shaded area) is largely due to strong \HH\ absorption lines. The inset shows spectra in this region obtained with different instruments (black line with blue shade), with the synthetic \HH\ profile convolved to the corresponding line spread function (red).\label{f:av}}
\end{figure*}

The overall spectral shape of \J\ appears to be typical of quasars (Fig.~\ref{f:av}).
 By matching the X-shooter data with the composite spectrum of \citet{Selsing2016}, shifted to the same emission redshift and reddened assuming  
a Small Magellanic Cloud type extinction law \citep{Gordon2003} at $z=4.1$, 
we inferred a moderate amount of dust in the DLA, with $\Av\sim0.06$~mag\footnote{The systematic uncertainty is comparable to the inferred value, due to intrinsic quasar shape variation; see 
\citet{Noterdaeme2017}.}.
In turn, bluewards of the quasar \lya\ emission, a very significant drop in transmission was observed. This results from the mutual blending of \HH\ lines in the DLA, as shown in the inset panel, together with unrelated intervening \HI\ absorption. 

Simultaneous multiple Voigt-profile fitting was then used to derive column densities ($N$), Doppler parameters ($b$), and redshifts ($z$) of the atomic and molecular absorption lines at $z=4.1$, including the effective atmospheric transmission in the total absorption model during the fit. Details on the analysis are provided in the Appendix, and we briefly summarise the obtained measurements here. 

Molecular hydrogen lines were detected across many rotational levels, with consistent features up to $J = 9$ (see Figs.~\ref{f:av} and \ref{f:H2J}). 
The column density was well constrained for the lowest rotational levels ($J<4$), which contain the bulk of H$_2$ and produce strongly damped lines. Furthermore, the damped lines are insensitive to the Doppler parameter and the velocity structure and are thus reasonably well constrained. From their excitation, we inferred a kinetic temperature of  
$T_k \approx T_{01}=97^{+15}_{-12}$~K.
Measuring the H$_2$ column densities in higher $J$-levels 
was very challenging due to the line blending (see Appendix), impeding their use to constrain the physical conditions.

From the metal and hydrogen column densities ($\log N(\HI)=20.8\pm0.1$; $\log N(\HH)=20.66\pm0.04$) in the system, we inferred an average gas-phase metal abundance of 
[Si/H]\,$\approx$\,-0.7. This metallicity is consistent with that expected from the $\sim$\,500~\kms\ wide velocity profile according to the metallicity--velocity relation observed by \citet{Ledoux2006dv90}. The overall molecular fraction was found to be $f_{\HH}= 2N(\HH)/(2N(\HH)+N(\HI)) = 0.59\pm 0.08$. 
This is a very conservative lower limit to the actual value in the \HH-bearing cloud since the observed $N(\HI)$ includes the many components seen in the metal-line profile. Assuming the overall metallicity is representative of that in the H$_2$-bearing component, we inferred a more realistic value of $f\approx0.9$ at $v=0$\,\kms, which is still conservative. Indeed, H$_2$ is expected to arise in the regions with the highest metallicity due to its formation on the surface of dust grains.  
Furthermore, \HI-\HH\ transition models \citep{Bialy2016} predict that beyond the dissociation front, the gas becomes fully molecular, with the observed \HI\ arising from the cloud's outer envelope. 
 Following these transition models, 
 the observed \HI\ column density and the moderate dust-to-gas ratio suggest a density-to-UV field ratio typical of cold gas with 
 $n/I_{\rm UV} > 100$\,cm$^{-3}$, where $I_{\rm UV}$ is relative to the \citet{Draine1978} interstellar radiation field (i.e. the mean Lyman-Werner band flux density is given by 
 $\overline{F}_{\nu} = 2.46\times10^{-8} I_{\rm UV}$\,photon\,cm$^{-2}$\,s$^{-1}$\,Hz$^{-1}$).

\section{Discussion and conclusions \label{s:discussion}}

So far, large molecular hydrogen column densities have been preferentially identified in systems pre-selected for their extreme \HI\ content. \citet{Ranjan2018} reported a very high \HH\ column density towards an extremely strong DLA \citep{Noterdaeme2014}, while \citet{Balashev2017} confirmed the presence of strong \HH\ absorption previously identified through an automated search from \citet{Balashev2014} applied to the SDSS-III DLA sample \citep{Noterdaeme2012sdss}. In these cases, the associated \HI\ column densities are among the highest known, 
in line with the scenario proposed by \citet{Noterdaeme2015b}, in which \HH\ is preferentially found in high-pressure environments, possibly located in the inner regions of galaxies. 
Supporting evidence includes the observed anti-correlation between the \HI\ column density and impact parameter \citep{Krogager2020a}, 
and the statistical detection of Ly-$\alpha$ emission superimposed on strong DLAs \citep{Noterdaeme2014}. We also note the detection of very large \HI\ and \HH\ columns in the afterglow spectrum of a $\gamma$-ray burst by \citet{Prochaska2009grb}. Since long-duration GRBs are linked to the deaths of massive stars, they likely trace dense, star-forming regions within galaxies \citep{Krogager2024}.

\begin{figure}[!ht]
    \centering
    \includegraphics[width=\hsize]{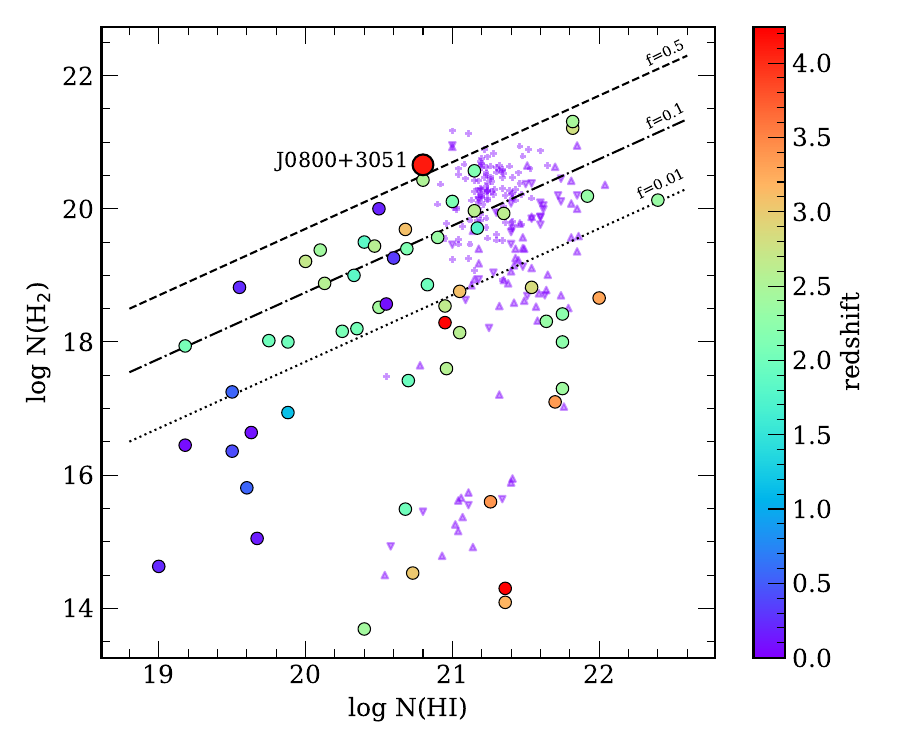}
    \caption{Column density of molecular vs atomic hydrogen measured from Lyman-Werner and Lyman lines, respectively. Measurements in intervening DLAs \citep[][and references therein]{Balashev2019} are represented by circles and coloured by redshift. The $z=4.1$ system towards \J\ is highlighted by a larger symbol. Measurements in the Local Group are depicted in purple (Milky Way: plus symbols by \citealt{Shull2021}; Small and Large Magellanic Clouds: upwards and downwards pointing triangles by \citealt{Welty2012}). The black lines indicate constant molecular fractions of 0.5, 0.1, and 0.01 (top to bottom).
    \label{fig:zf}}
\end{figure}

In this work, not only does the DLA have the highest molecular fraction directly measured to date at $z>0$, but it is also found among the very highest redshifts so far (Fig.~\ref{fig:zf}), making its detection surprising by itself. The GMOS survey by \citet{Crighton2015}, which includes the object presented in this paper, contains 163 quasars with 45 DLAs at $z>4$. For comparison, only 
two intervening absorbers with $\log N(\HH)>20.5$ have been identified among $\sim$10$^4$ DLAs in SDSS at lower redshifts. 
Moreover, if the \HH\ column density distribution, inferred at $z\sim2.5$ by \citet{Krogager2020} 
from the stacking analysis of \citet{Balashev2018}, applies at $z\sim4$, 
only $\sim 0.1$ systems with $\log N(\HH) > 20.5$ would be expected in the GMOS survey.
The key question is then whether such a difference in detection rate reflects a genuine physical evolution or a combination of selection effects.

From a physical perspective, higher mean gas densities at high redshift, as proposed by \citet{Stern2021}, could in principle enhance \HH\ formation. In addition, more turbulent neutral gas driven by stronger energy injection from star formation \citep[e.g.][]{Green2010}, accretion \citep[e.g.][]{Elmegreen2010}, and mergers \citep[e.g.][]{Bournaud2011} could promote cold gas and \HH\ formation through local density enhancements \citep[see also][]{Colman2025}. These effects may dominate over the opposing influence of lower average metallicities and a stronger ultraviolet radiation field.

On the other hand, our detection may highlight observational selection effects that enter at multiple stages of the detection process.
The first step of the detection process
involves the identification of the background quasar. In previous quasar surveys, this process is based on a set of complex colour selection criteria and partly builds how the intergalactic medium transmission affects the quasar's spectrum. 
This colour-based selection 
is notoriously difficult at $z\sim2.5-3$ due to the quasar colours crossing the stellar locus \citep{Richards2002,Krogager2019}.
Strong \HH\ lines can further absorb a significant fraction of the quasar light bluewards of about 110\,nm in the absorber's rest frame, making the quasar identification even more difficult. In the present high-redshift case, the absorber only contributes to the $u,g$ drop-out, which is central to the selection of high-redshift quasars and does not alter the colours much, especially since its dust content remains modest.
However, in more dust-reddened systems, colour-based selection may become inefficient even at these redshifts. This highlights the importance of complementary selection strategies that are less sensitive to colours \citep[e.g.][]{Krogager2018,Krogager2023}.

Once a spectrum of the quasar has been obtained, the second challenge is the detection of the absorption system itself, which the present system perfectly illustrates by being unnoticed for two decades in public survey data.
An efficient strategy to identify H$_2$ systems is to search for \CI\ absorption, which has been shown to be an excellent tracer of molecular gas \citep{Noterdaeme2018}.  Strong transitions are available at 1560 and 1656\,\AA, i.e. redwards of the Ly-$\alpha$ forest, which facilitates their detection \citep[e.g.][]{Ledoux2015}. Nevertheless, \CI\ absorbers tend to select high-metallicity and dusty absorbers, and they may miss low metallicity molecular systems. 
Correlation techniques based on \HH\ line templates provide a particularly powerful method to recognise the absorption pattern, even when little transmission remains between the \HH\ bands. Such methods also have the advantage of being insensitive to the properties of the metals or \HI\ absorption. In fact, the detection presented here arose as a by-product of testing such detection algorithms for \HH\ absorption at the quasar redshift \citep{Noterdaeme2019}.

By comparing new observations with archival data, we are now able to examine how the same \HH\ absorber manifests itself across surveys with different spectral resolutions, wavelength coverage, and signal-to-noise ratios (see Fig.~\ref{f:av}). This provides a rare empirical benchmark for assessing and refining selection and detection algorithms, and it is arguably more informative than mock spectra, as it relies on a real astrophysical system.

Finally, even when a strong \HH\ absorber is identified, obtaining robust physical constraints remains challenging. As in the present case, the total column density and kinetic temperature are primarily constrained by the low rotational levels, provided that some transmission remains between absorption bands. 
The reddest Lyman bands of \HH\ are particularly valuable, as they are less affected by mutual blending, whereas lines at shorter wavelengths merge into a quasi-continuum absorption similar to a Lyman limit. As a result, the number of usable transitions decreases at high column densities, making a high S/N and high spectral resolution essential to reliably constraining the rotational populations and inferring physical conditions.

Overall, detecting and characterising extremely strong \HH\  absorbers remains challenging. The detection presented in this work may at first glance argue against a significant bias. After all, we did manage to identify this absorption system 
despite expecting only $\sim$0.1.
In turn, this means that we were extremely lucky, the formation of \HH\ is more efficient at early cosmic times, 
or we are missing substantially more \HH\ at lower redshifts than currently anticipated.
The low dust reddening of this system may have also been crucial to its detection, suggesting that dustier systems are still being missed, which would strengthen these points even more.

Further progress at all redshifts, possibly pushing towards even more extreme and likely rarer systems, will require both more complete quasar selection strategies and further development of efficient \HH\ detection techniques. 
While such absorbers offer a unique window into the dense phases of the neutral gas at high-$z$, our limited ability to constrain the physical conditions and environment, even when detected, highlights the need for next-generation facilities. Thirty-metre class telescopes equipped with high resolution spectrographs, combined with deep emission observations aimed at detecting the associated galaxies, will be crucial to fully characterising these systems.

\begin{acknowledgements}
We thank the referee for useful comments and suggestions.
We thank ESO for granting Director’s Discretionary Time and Paranal staff for executing our observations. We thank Rodrigo Cuellar for modelling the atmospheric transmission using Molecfit. JKK acknowledges financial support through the French Agence Nationale de la Recherche (ANR) under grant number ANR-24-CE31-7454 (CI-CNM). SB thanks IAP for hospitality and support 
during a science visit.
\end{acknowledgements}


\bibliographystyle{aa} 

\begin{appendix}

\section{Notes on absorption modelling and column density measurements}

\subsection{Atomic hydrogen}

\begin{figure}[!ht]
    \centering
    \includegraphics[trim={0.5cm 0.5cm 0.1cm 0.1cm},width=\hsize,clip=]{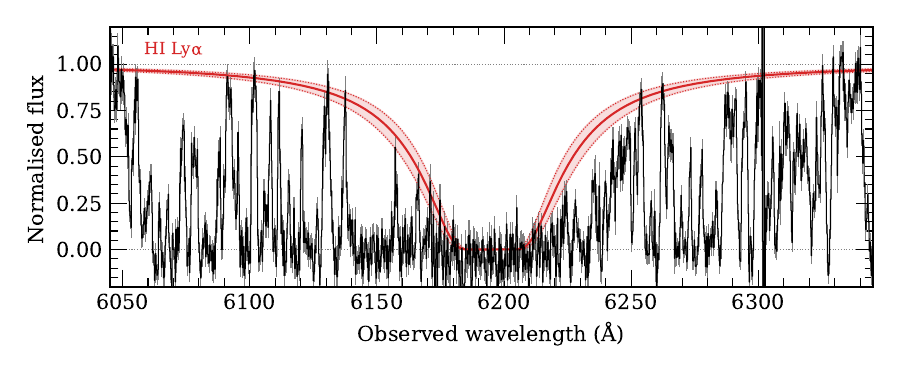}
    \caption{Damped \HI\ \lya\ profile. The data is shown in black with uncertainties in grey. The model with $\log N(\HI)=20.8$ is shown in red with 0.1~dex range shown by the shaded area.}
    \label{f:HI}
\end{figure}

The total atomic hydrogen column density in the DLA was constrained from the \lya\ line. Because of contaminating absorption from \HI\ forest in the wings, the fit was constrained using the local transmission peaks, yielding $\log N(\HI)=20.8 \pm 0.1$, see Fig.~\ref{f:HI}. This value matches the estimate based on SDSS data by \citet{Prochaska2005sdss} but is lower than that from GMOS data \citep[$\log N(\HI)=21.05$;][]{Crighton2015} probably due to contamination by the \lya\ forest unrecognised at the lower achieved spectral resolution.

\subsection{Low-ionisation metal lines}

\begin{figure}
    \centering
    \includegraphics[width=\hsize]{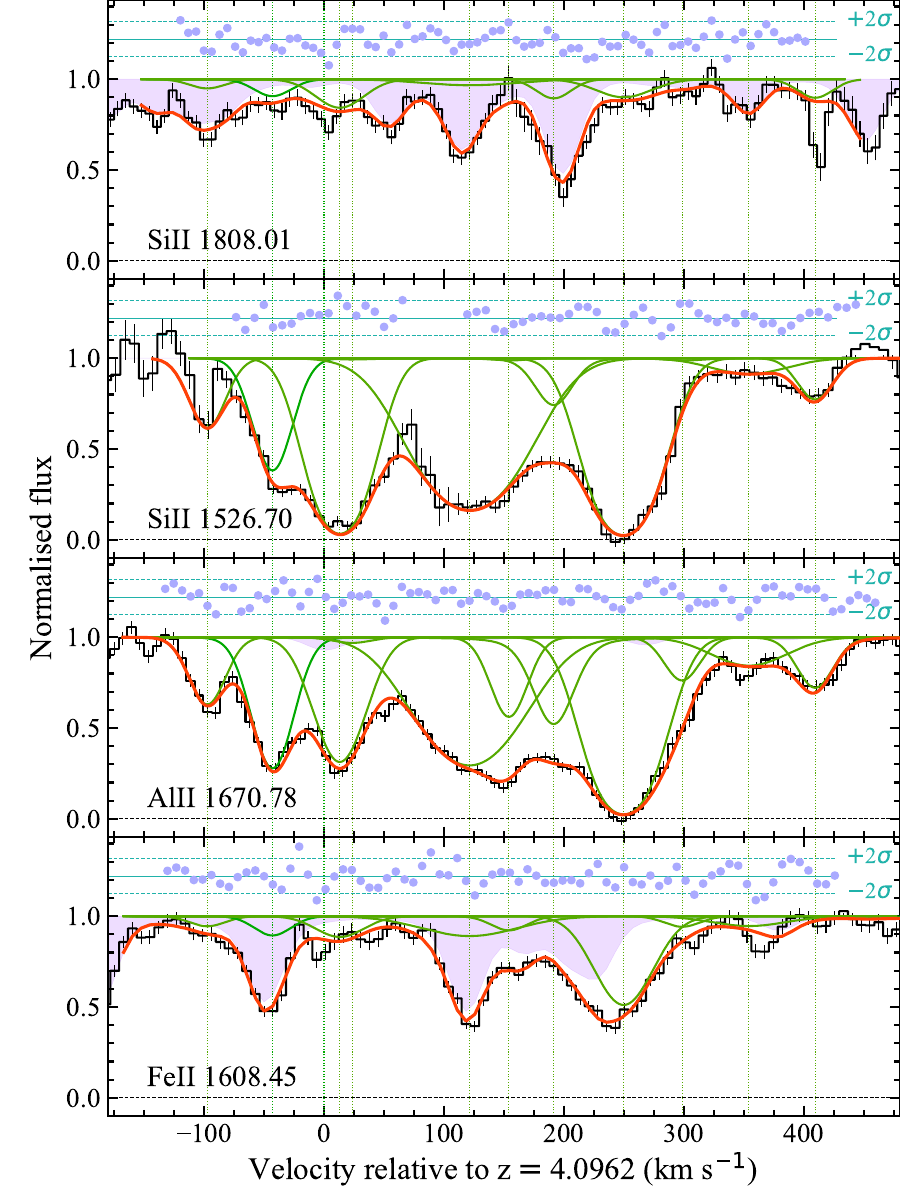}
    \caption{Low-ionisation metal absorption lines. The data is shown in black with the overall absorption model in red. Contribution from individual components is shown in green and that from telluric lines depicted as purple shaded area. Note that two regions in the \SiII$\lambda$1526 profile at about $-100$ and 
    $+100$~\kms\ are affected by sky emission line residuals and not used to constrain the fit. Residuals are shown in the top panels.}
    \label{fig:placeholder}
\end{figure}

Metals
absorption lines from low-ionisation species (\SiII, \FeII, and \AlII) are detected in multiple components spanning approximately 500~\kms. Owing to the high redshift, only \SiII$\lambda1526$, \SiII$\lambda1808$, \FeII$\lambda1608$, and \AlII$\lambda1670$ fall redwards of the \lya\ forest. Among these, \SiII$\lambda1808$ and \FeII$\lambda1608$ are weak and partially blended with telluric absorption, while the other two transitions contain components in the intermediate regime. Consequently, although a simultaneous fit assuming common redshift and Doppler parameter ($z, b$) across species provides a visually good fit to the data, the column densities in the individual components could be degenerate. The total column densities,  provided in Table~\ref{t:colden}, should however be more robust. 

\subsection{Neutral carbon}

\begin{figure}
    \centering
    \includegraphics[trim={0.cm 1cm 0.0cm 0.0cm},width=\hsize,clip=]{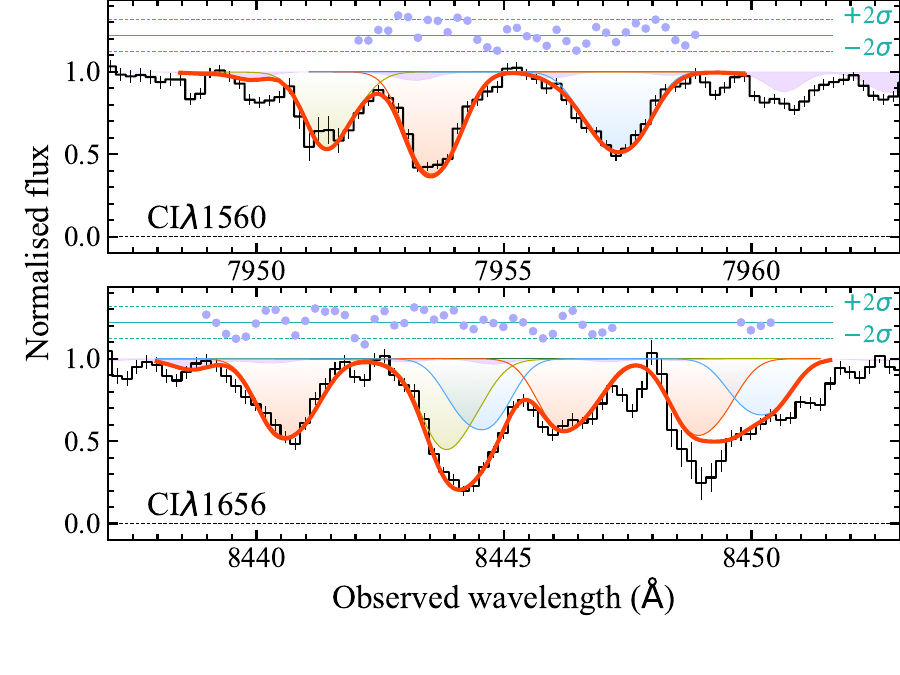}
    \caption{\CI\ absorption lines. The data are shown in black with the overall absorption model in red. Contribution from the different fine-structure levels are shown by green (J=0), orange (J=1) and blue (J=2) shaded areas while that of telluric lines is shown in purple.}
    \label{f:CI}
\end{figure}

Neutral carbon absorption lines are clearly detected at the velocity of the \HH\ component along with the three fine-structure levels. A slight asymmetry suggests the presence of two components, which we used to model the $\lambda$1560 and 1656 lines that are located outside the \lya\ forest, see Fig.~\ref{f:CI}. Intrinsic saturation leads to large uncertainties on their column densities as reported in Table~\ref{t:colden}. 

\begin{table}[]
    \centering
        \caption{Total column densities measured in the $\zabs=4.1$ DLA towards \J. \label{t:colden}}
    \begin{tabular}{c c}
    \hline \hline
        Species & $\log N/\cmsq$ \\
        \hline
        \HI     &    20.8 $\pm$ 0.1             \\
        \SiII   &  $15.66^{+0.34}_{-0.10}$ \\
        \FeII   &  $14.50\pm 0.03$              \\
        \AlII   &  $15.09^{+0.15}_{-0.13}$              \\
        \CI,J=0 &  $16.5^{+0.4}_{-0.8}$         \\
        \CI,J=1 &  $15.2^{+0.8}_{-0.3}$         \\
        \CI,J=2 &  $14.58^{+0.08}_{-0.08}$      \\
        \HH,J=0 &  20.18 $\pm$ 0.10             \\
        \HH,J=1 &  20.37 $\pm$ 0.03        \\
        \HH,J=2 &  19.50 $\pm$ 0.20             \\
        \HH,J=3 &  19.46 $\pm$ 0.15             \\
        \HH,J=4 &  18.2-19.2              \\
        \HH,J=5 &  18.0-19.2              \\
        \HH,J=6 &  16.6-18.8              \\
        \HH,J=7 &  16.1-18.2              \\
        \HH,J=8 &  15.1-15.9              \\
        \HH,J=9 &  15.1-16.0              \\
        total \HH &  20.66 $\pm$ 0.04\,\tablefootmark{$\dagger$}             \\
         \hline
    \end{tabular}
    \tablefoot{\tablefoottext{$\dagger$}{Obtained from summing the central values across levels. The quoted uncertainty represents the formal propagation of the model ranges, treating them as 1\,$\sigma$ standard errors.}}
    \label{tab:placeholder}
\end{table}

\subsection{Modelling the H$_2$ absorption profile}

\begin{figure*}
    \centering
    \includegraphics[trim={0cm 0.3cm 0cm 0cm},width=\hsize,clip=]{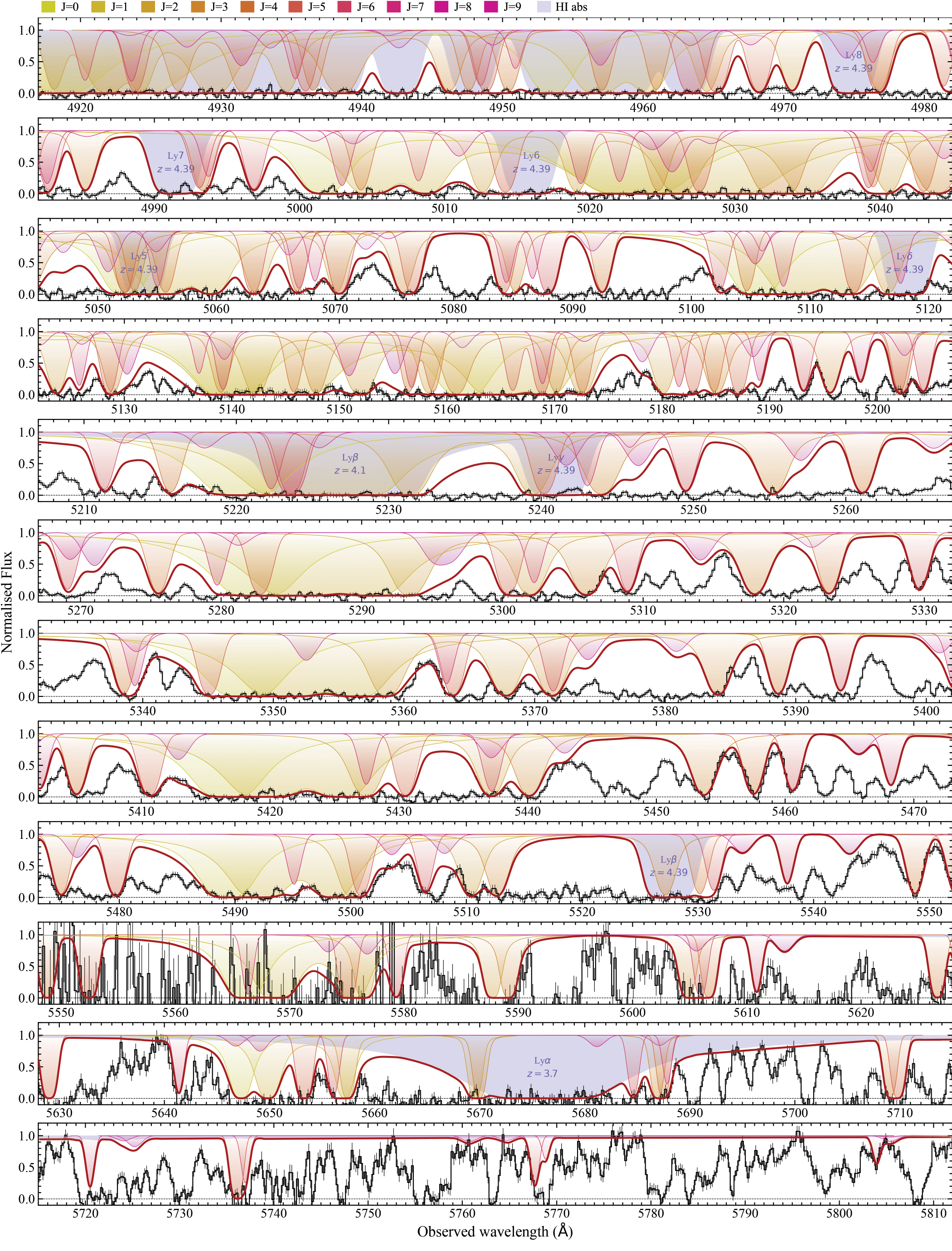}
    \caption{Portion of the normalised X-shooter spectrum (black) with the overall fitted absorption model (red line). The contribution from low to high rotational levels of H$_2$ at $z=4.0962$ is shown by the yellow to red gradient fill (see top legend). 
    Blue areas represent \HI\ absorption from the $z=4.096$ DLA and the sub-DLAs at $z=3.669$ and $z=4.387$.  
    }
    \label{f:H2J}
\end{figure*}

Because of the strong blending among the \HH\ lines, further affected by contamination from the \HI\ forest, constraining the \HH\ column densities is challenging. We found that different modelling choices and assumptions can yield significantly different column densities in the high rotational levels while providing comparably good fits to the data. We therefore adopted a conservative approach and report the full range of column densities obtained from a variety of models: (i) a single-component H$_2$ model with a common Doppler parameter for all rotational levels; (ii) the same model allowing the Doppler parameter to increase with rotational level, as sometimes observed in the literature \citep[e.g.][]{Lacour2005}; and (iii) two-component models, motivated by a possible asymmetry in the high-$J$ lines and in the \CI\ profiles, with either tied or level-dependent Doppler parameters. 
We also considered the cumulative absorption from the \HI\ forest as a pseudo-continuum variation. Since the dense and heavily blended \HI\ lines make the detailed transmission difficult to model, their net effect was approximated by a linear correction to the unabsorbed quasar continuum. 
The inferred column densities are highly consistent across models up to $J=3$, for which we give the central point with range as uncertainty but diverge substantially for $J\geq4$, for which we provide a range only in Table~\ref{t:colden}. The contribution of lines from different rotational levels to the overall absorption model is highlighted in Fig.~\ref{f:H2J}.

\subsection{Additional high column density \HI\ systems}

We also identified a strong sub-DLA at $z=3.669$ with $\log N(\HI)\simeq 20$, accompanied by low-ionisation metal absorption lines (\SiII$\lambda$1526, \AlII$\lambda$1670). In addition, we derived a slightly revised absorption redshift ($z=4.387$ instead of $z=4.386$) based on the \SiIV\ and \CIV\ features observed in the X-shooter data for the Lyman-limit system reported by \citet{Crighton2019}. These systems are included in the overall absorption model shown in Fig.~\ref{f:H2J}.

\end{appendix}

\end{document}